\documentclass[10pt]{article}

\usepackage{amsmath}
\usepackage{amssymb}

\usepackage{graphicx}

\usepackage{cite}

\usepackage{color} 

\usepackage[labelfont=bf,labelsep=period,justification=raggedright]{caption}

\makeatletter
\renewcommand{\@biblabel}[1]{\quad#1.}
\makeatother

\date{}

\begin{document}

\begin{flushleft}
{\Large
\textbf{The Strength of Friendship Ties in Proximity Sensor Data}
}
\\
Vedran Sekara$^{1,\ast}$,  
Sune Lehmann$^{1,2}$
\\
\bf{1} Cognitive Systems, Department of Applied Mathematics and Computer Science, Technical University of Denmark, Kgs.~Lyngby, Denmark\\
\bf{2} Niels Bohr Institute, University of Copenhagen, \O{}sterbro, Denmark
\\
$\ast$ E-mail: Corresponding vese@dtu.dk
\end{flushleft}

\section*{Abstract}
Understanding how people interact and socialize is important in many contexts from disease control to urban planning. Datasets that capture this specific aspect of human life have increased in size and availability over the last few years. We have yet to understand, however, to what extent such electronic datasets may serve as a valid proxy for real life social interactions. For an observational dataset, gathered using mobile phones, we analyze the problem of identifying transient and non-important links, as well as how to highlight important social interactions. Applying the Bluetooth signal strength parameter to distinguish between observations, we demonstrate that weak links, compared to strong links, have a lower probability of being observed at later times, while such links---on average---also have lower link-weights and probability of sharing an online friendship. Further, the role of link-strength is investigated in relation to social network properties.

\section*{Introduction}
Recognizing genuine social connections is a central issue within multiple disciplines. When do connections happen? Where do they take place? And with whom is an individual connected? These questions are important when working to understand and design urban areas \cite{sun2011exploring, sevtsuk2010does}, studying close-contact spreading of infectious diseases \cite{liljeros2001web, mossong2008social, cauchemez2009household}, or organizing teams of knowledge workers \cite{wu2008mining, pentland2012new, blansky2013spread}.
In spite of their importance, measuring social ties in the real world can be difficult.

In classical social science the standard approach is to use self-reported data. This method, however, is only practical for relatively small groups and suffers from cognitive biases, errors of perception, and ambiguities \cite{wuchty2009social}. Further, it has been shown that the ability to capture behavioral patterns via self-report data is limited in many contexts \cite{watts2007twenty}. A different approach for uncovering social behavior is to use digital records from emails and cell phone communication \cite{eckmann2004entropy, barabasi2005origin, kossinets2006empirical, onnela2007structure, gonzalez2008understanding, lazer2009computational, song2010limits, bagrow2011collective}. Although such analyses have improved our understanding of social ties, they have left many important questions unanswered---are electronic traces a valid proxy for real social connections? 
Eagle et al.~\cite{eagle2009inferring} began to answer this question by including a spatial component as part of their data, using the short range ($\sim 10\,m$) Bluetooth sensor embedded in study participants' smartphones to measure physical proximity. Their results show that proximity data closely reflects social interactions in many cases. But since it is easy to think of examples where reciprocal Bluetooth detection does not correspond to social interaction (e.g.~transient co-location in dining hall) the question remains, which observations correspond to actual social interactions and which are just noise?

Multiple alternatives have been proposed to Bluetooth for sensor-driven measurement of social interactions, each with particular strengths and weaknesses \cite{haritaoglu2000w,polastre2005telos, salathe2010high, rosenstein2008video, olguin2009sensible, kjaergaard2012challenges, wyatt2007capturing, carreras2012comm2sense, wang2011human, cattuto2010dynamics, barrat2013temporal}. For example, Radio Frequency Identification (RFID) badges have short interaction ranges ($1-4\,m$) and measure only face-to-face interactions, thus solving many of the resolution problems posed by Bluetooth \cite{cattuto2010dynamics, barrat2013temporal}. This approach, however, confines interactions to occur within specific areas covered by special radio receivers and requires participants to wear custom radio tags on their chests at all times---unlike Bluetooth which is ubiquitous across many types of modern electronic devices. 

Our investigation digs into the role of Bluetooth signal strength, using a dataset obtained from applications running on the cell phones of 134 students at a large academic institution. Each phone records and sends data to researchers about call and text logs, Bluetooth devices in nearby proximity, WiFi hotspots in proximity, cell towers, GPS location, and battery usage\cite{stopczynski2014measuring}. In addition, we combine the data collected via the phones with online data, such as social graphs from Facebook for a majority of the participants. The study continuously gathers data, but in this paper we focus on Bluetooth proximity data gathered for 119 days during the academic year of 2012-2013. Specifically, we focus on the received signal strength parameter and propose a methodology that applies signal strength to distinguish between social and non-social interactions. We concentrate on the signal parameter because it is present in a majority of digitally recorded proximity datasets \cite{ladd2005robotics,cattuto2010dynamics,stopczynski2014measuring} and in addition, it also suggests a rough estimate for the distance between two devices.
Applying the method on our data, we compare the findings to a null model and demonstrate how removing links with low signal strength influences network structure. Moreover, we use estimated link-weights and an online dataset to validate the friendship-quality of removed links.

\section*{Materials and Methods}
\subsection*{Dataset}
We distributed phones among students from four study lines (majors), where each major was chosen based on the fraction of students interested in participating in the project. This selection method yielded a coverage of $>93 \%$ of students per study line, enabling us to capture a dense sample of the social interactions between subjects.
Such high coverage of internal connections within a social group, with respect to the density of social interactions combined with the duration of observation, has not been achieved in earlier studies \cite{eagle2009inferring,cattuto2010dynamics}.

The data collector application installed on each phone follows a predefined scanning time table, which specifies the activation and duration of each probe. Proximity data is obtained by using the Bluetooth probe. Every 300 seconds each phone performs a Bluetooth scan that lasts 30 seconds. During the scan it registers all discoverable devices within its vicinity ($5-10 m$) along with the associated received signal strength indicator (RSSI) \cite{shorey2000bluetooth}.  Recorded proximity data is of the form ($i$, $j$, $t$, $s$), denoting that person $i$ has observed $j$ at time $t$ with signal strength $s$. Only links between experiment participants are considered, comprising a dataset of $2\,183\,434$ time ordered edges between $134$ nodes, see Table \ref{tab:data} for more information.
Data collection, anonymization, and storage was approved by the Danish Data Protection Agency, and complies with both local and EU regulations. Written informed consent was obtained via electronic means, where all invited participants digitally signed the form with their university credentials. 
Along with the mobile phone study we also collected Facebook graphs of the participants. Not all users donated their data since this was voluntary, however we obtained a user participation of $\sim88\%$ (119 users and 1018 Facebook friendships). For the missing $12\%$ of users, we assume they do not share any online friendships with the bulk of participants.

\begin{table}[!htbp]
\begin{center}

\begin{tabular}{|c|c|c|}
\hline
& Total & Average pr. time-bin \\
\hline
Nodes (Users) & $134$ & 17.32 \\
\hline
Edges (Dyads) & $2\,183\,434$ & 62.50\\
\hline
Time-bins & $34\,272$ & - \\
\hline
Average clustering & 0.85 & 0.26 \\
\hline
Average degree & 103.51 & 2.41 \\
\hline
\end{tabular}
\end{center}
\caption{
{\bf{Data overview}} \small Statistics showing the number of total (aggregated) and average values of network properties. Time-bins span five minutes and cover the entire 119 day period, including weekends and holidays. For the average values we only take active nodes into account, i.e. people that have observed another person or been observed themselves in that specific time-bin. Network properties are calculated for the full aggregated network and as averages over each temporal network slice.
}
\label{tab:data}
\end{table}

\subsection*{Identifying links}
Independent of starting conditions, the scanning framework on one phone will drift out of sync with the framework on other phones after a certain amount of time, thus the phones will inevitably scan in a desynchronized manner. This desynchronization can mainly be attributed to: internal drift in the time-protocol of each phone, depletion of the battery, and users manually turning phones off.
To account for irregular scans, we divide time into windows (bins) of fixed width and aggregate the Bluetooth observations within each time-window into a weighted adjacency matrix. 
The complete adjacency matrix is thus given by: $W=\left(W^{(\Delta t_1)},W^{(\Delta t_2)},\dots,W^{(\Delta t_n)}\right)$, where each link is weighted by its signal strength and where $\Delta t_i$ indicates window number $i$. These matrices generally assume a non-symmetric form, i.e. person $A$ might observe $B$ with signal strength $s$ while person $B$ observes $A$ with strength $s'$, or not at all. 
The scanning frequency of the application sets a natural lower limit of the network resolution to $5$ minutes. 
If we are interested in the social dynamics at a different temporal resolution we can aggregate the adjacency matrices and retain entries according to some heuristic (e.g. with the strongest signal).  Depending on the level of description (monthly, weekly, daily, hourly, or every 5 minutes) the researcher must think carefully about the definition of a network connection. Frameworks for finding the best temporal resolution, so called \textit{natural timescales} have for specific problems been investigated by Clauset and Eagle \cite{clauset2012persistence}, and Sulo et al. \cite{sulo2010meaningful}. In this paper, however, we are interested in the identification and removal of non-important proximity links, so aggregating multiple time-windows is not a concern here. Henceforth we solely work with 5 minutes time-bins.

The Bluetooth probe logs all discoverable devices within a sphere with a radius of 5-10 meters---walls and floor divisions reduce the radius, but the reduction in signal depends on the construction materials \cite{cheung2006inexpensive}. Blindly taking proximity observations as a ground truth for social interactions will introduce both false negative and false positive links in the social network. False negative links are typically induced by hardware errors beyond our control, thus we focus on identifying false positive links. We therefore propose to identify non-social or noisy proximity links via the signal strength parameter. The parameter can be thought of as a proxy for the relative distance between devices, since most people carry their phones on them, it will in principle also suggests the separation distance between individuals.

Previous work has applied Bluetooth signals to estimate the position of individuals \cite{anastasi2003experimenting, bruno2003design, madhavapeddy2005study, zhou2006position} but studies by Hay \cite{hay2009bluetooth}, and Hossein et al.~\cite{hossain2007comprehensive} have revealed signal strength as an unsuitable candidate for accurately estimating location. However, the complexity of the problem can greatly be reduced by focusing on the relative distance between individuals rather than position. 
In theory, the transmitted power between two antennae is inversely proportional to the distance squared between them \cite{friis1946note}. Reality is more complicated, due to noise and reflection caused by obstacles. 

\begin{figure}[!htpb]
\begin{center}
\includegraphics[width=0.9\linewidth]{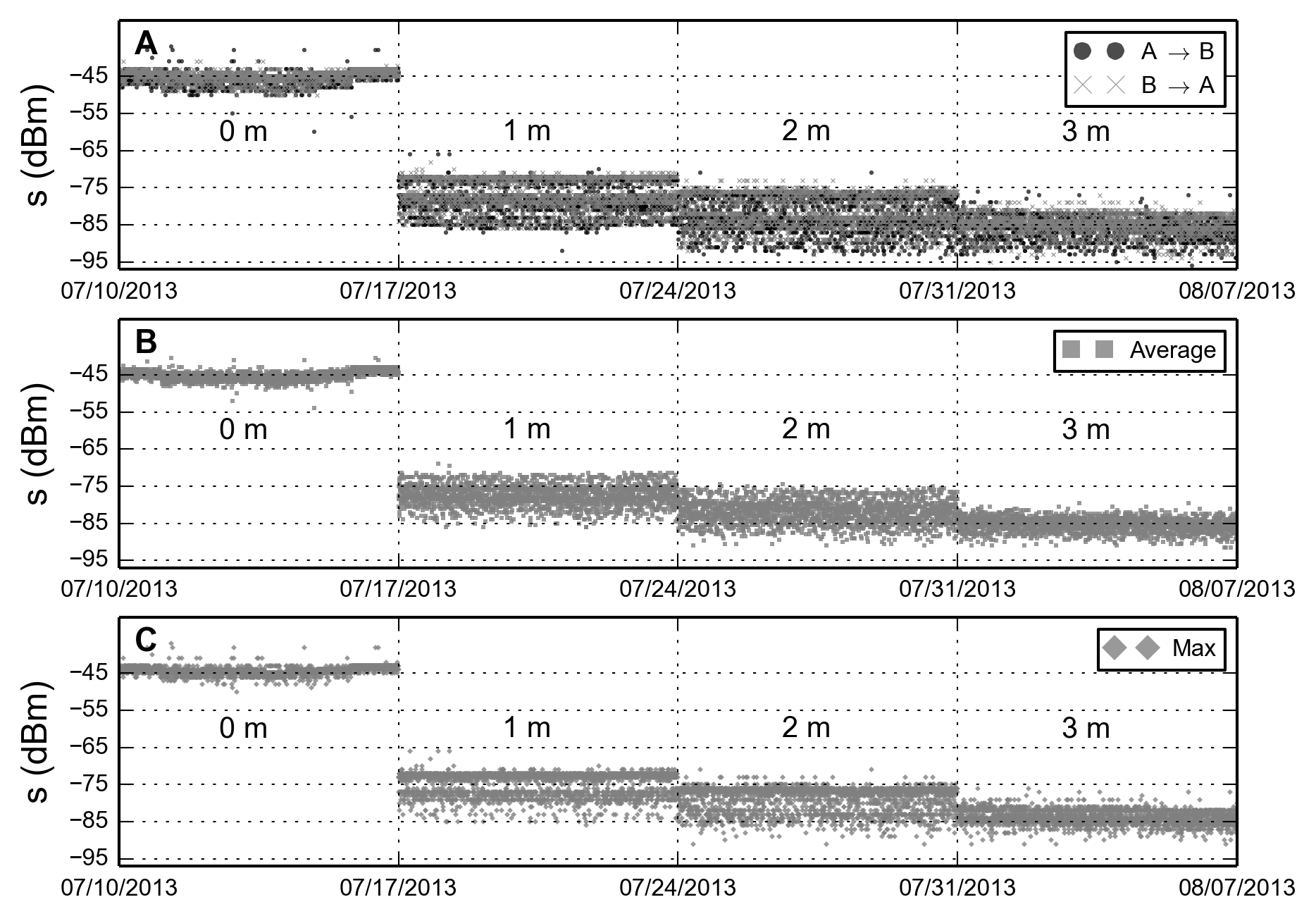}
\end{center}
\caption{ \small
{\bf Bluetooth signal strength (RSSI) as a function of distance.} \textbf{A:} Scans between two phones. Measurements are per distance performed every five minutes over the course of 7 days. Mean value and standard deviation per distance are respectively $\mu_0=-45.13\pm1.56\text{ dBm}$, $\mu_1=-77.48\pm4.15\text{ dBm}$, $\mu_2=-82.03\pm4.57\text{ dBm}$, and $\mu_3=-85.49\pm2.75\text{ dBm}$. \textbf{B:} Average of the values in respective time-bins. Summary statistics are: $\mu_0^{\text{avg}}=-45.13\pm1.20\text{ dBm}$, $\mu_1^{\text{avg}}=-77.46\pm2.90\text{ dBm}$, $\mu_2^{\text{avg}}=-81.99\pm3.17\text{ dBm}$, and $\mu_3^{\text{avg}}=-85.45\pm1.88\text{ dBm}$.  \textbf{C:} Maximal value per time-bin. The mean value and standard deviation per distance are: $\mu_0^{\text{max}}=-44.41\pm1.11\text{ dBm}$, $\mu_1^{\text{max}}=-75.09\pm3.24\text{ dBm}$, $\mu_2^{\text{max}}=-79.25\pm3.47\text{ dBm}$, and $\mu_3^{\text{max}}=-83.88\pm2.00\text{ dBm}$. The measurements cover hypothetical situations where individuals are far from each other and on either side of a wall.
}
\label{fig:distance}
\end{figure}

We use the ideal result as a reference while we perform empirical measurements to determine how signal strength depends on distance. Two devices are placed on the ground in a simulated classroom setting, where we are able to control the relative distance between them.
The resulting measurements are plotted in Fig.\,\ref{fig:distance}A. As is evident from the figure, there is a large variance in the measured signal strength values for each fixed distance. However, as both phones exhibit the same variance we can exclude faulty hardware; further, environmental noise such as interference from other devices, or solar radiation can also be dismissed since there appear no daily patterns in the data. But we observe multiple bands or so-called modes onto which measurements collapse, Ladd et al.~\cite{ladd2005robotics} noted a similar behavior for the received signal strength of WiFi connections, both are phenomena caused by non-Gaussian distributed noise.
The empirical measurements form a foundation for understanding signal variance as a function of distance, but they were performed in a controlled environment. In reality, there are a multitude of ways to carry a smartphone: some carry it around in a pocket, others in a bag. Liu and Striegel \cite{liu2011accurate} investigated how these various scenarios influence the received signal strength---their results indicate only minor variations, hence we conclude that the general behavior is similar to the measurements shown in the figure. Further, social interactions are not only limited to office environments, so we have re-produced the experiment in outdoors and in basement-like settings; the results are similar.

\begin{figure}[!htbp]
\begin{center}
\includegraphics[width=0.9\linewidth]{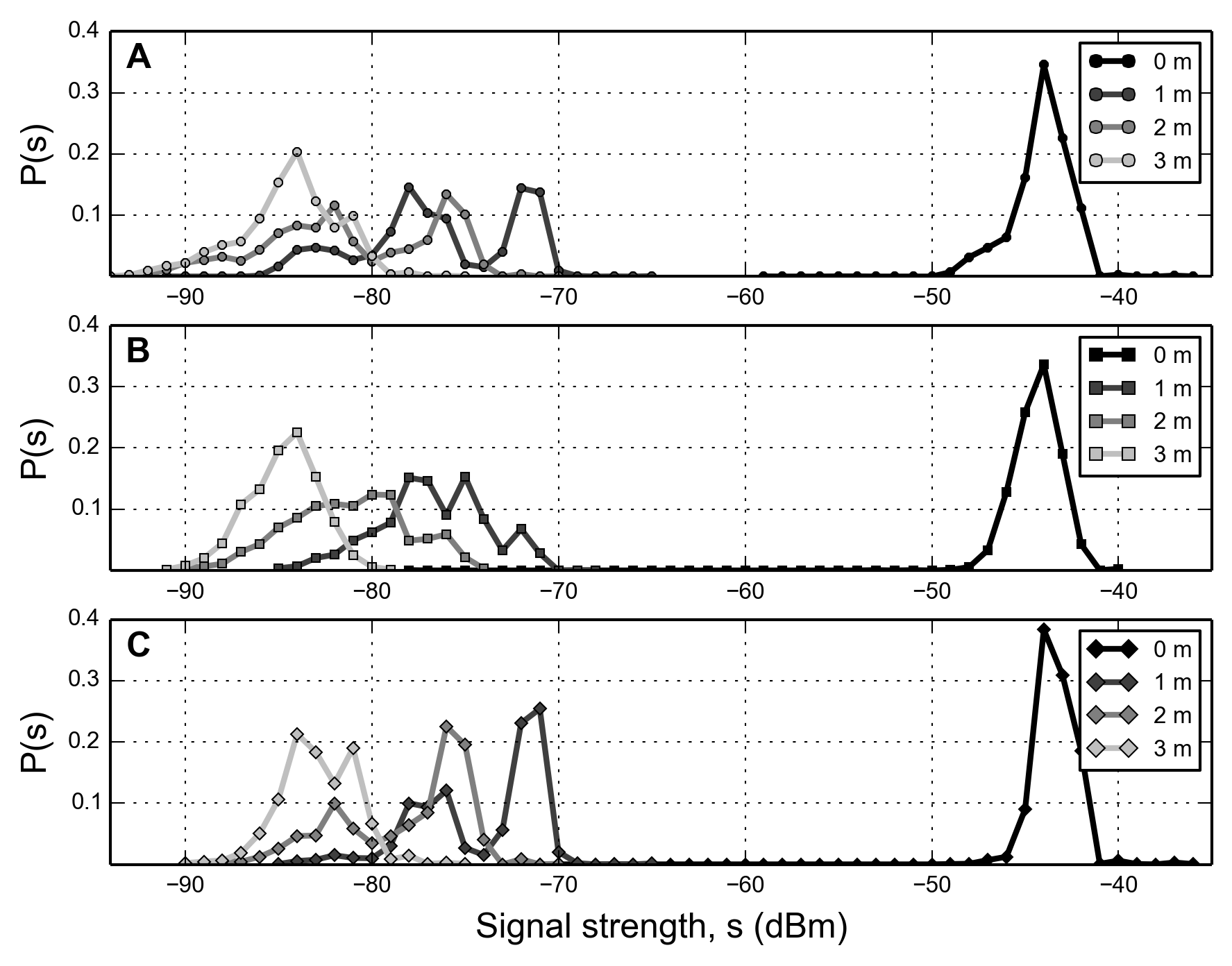}
\end{center}
\caption{ \small
{\bf Distributions of signal strength for the respective distances.}  \textbf{A:} Raw data. Measurements from both phones are statistically indistinguishable and are collapsed into single distributions, i.e. there is no difference between whether $A$ observes $B$ or vise versa. \textbf{B:} Average of signal strength per time-bin. \textbf{C:} Maximal value of signal strength per. time-bin.
}
\label{fig:distance_dist}
\end{figure}

Bi-directional observations yield at most two observations per dyad per 5-minute time-bin, we can average over the measurements (Fig\,\ref{fig:distance}B), or take the maximal value (Fig \ref{fig:distance}C). Fig.\,\ref{fig:distance_dist} shows the distributions of signal strength for each respective distance. For raw data, Fig.\,\ref{fig:distance_dist}A, we observe a localized zero-distance distribution while the 1, 2, and 3-m distributions overlap considerably. Averaging over values per time-bin smoothes out and compresses the distributions, but the bulk of the distributions still overlap (Fig. \ref{fig:distance_dist}B). Taking only the maximal signal value into account separates the distributions more effectively (Fig.\,\ref{fig:distance_dist}C). The reasoning behind choosing the maximal signal value is that phones are physically at different locations and we expect the distance to be maximally reflected in the distributions. 

\begin{figure}[!htbp]
\begin{center}
\includegraphics[width=0.4\linewidth]{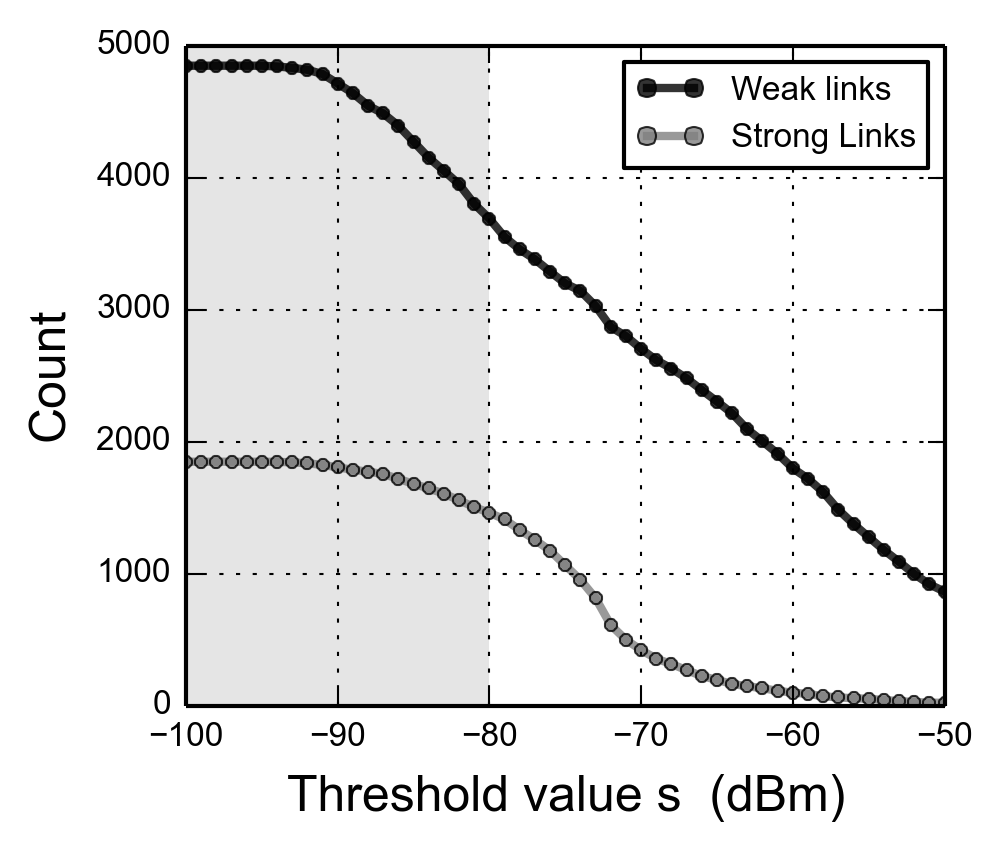}
\end{center}
\caption{ \small
{\bf Number of links per type as a function of threshold value.}  Links are classified as weak if they are observed less than $120$ times in the data, i.e. links that on average are observed less than once per day---otherwise they are classified as strong. Grouping students into study lines, reveals that links within each study line have an almost uniform distribution of weights while links across study lines are distributed according to a heavy-tailed distribution. A threshold of $-80$ $dBm$ (gray area) removes 1159 weak and 387 strong links and classifies $97.6 \%$ of inter-study line links as weak and $86.7 \%$ of intra-study line links as strong.
}
\label{fig:threshold_curve}
\end{figure}

Thus, by thresholding observations on signal strength, we can filter out proximity links that are likely to be further away than a certain distance. By doing so we are able to emphasize links that are more probable of being genuine social interactions, while minimizing noise and filtering away non-social proximity links. From the behavioral data we count the number of appearances per dyad and assign the values as weights for each link. 
Link weights follow a heavy-tailed distribution, with a majority of pairs only observed a few times (low weights), a social behavior that has previously been observed by Onnela et al. \cite{onnela2007structure}. 
Based on their weight we divide links into two categories: weak and strong. A link is defined as `weak' if it has been observed (on average) less than once per day during the data collection period, remaining links are characterized as `strong'. An effective threshold should maximize the number of removed weak links, while minimizing the loss of strong links. Fig. \ref{fig:threshold_curve} depicts the number of weak and strong links as a function of threshold value. We observe that, as we increase the threshold, the number of weak links decreases linearly, while the number of strong links remains roughly constant and then drops off suddenly. 
Taking into account both the maximum-value distance distributions (Fig. \ref{fig:distance_dist}C) and link weights (Fig.\,\ref{fig:threshold_curve}), we choose the value ($-80 \,dBm$) that optimizes the ratio between strong and weak links. In a large majority of cases, this corresponds to interactions that occur within a radius of $0 - 2$ meters---a distance which Hall \cite{hall1990hidden} notes as a typical social distance for interactions among close acquaintances.

\subsection*{Removing links}
\begin{figure}[!htbp]
\begin{center}
\includegraphics[width=0.8\linewidth]{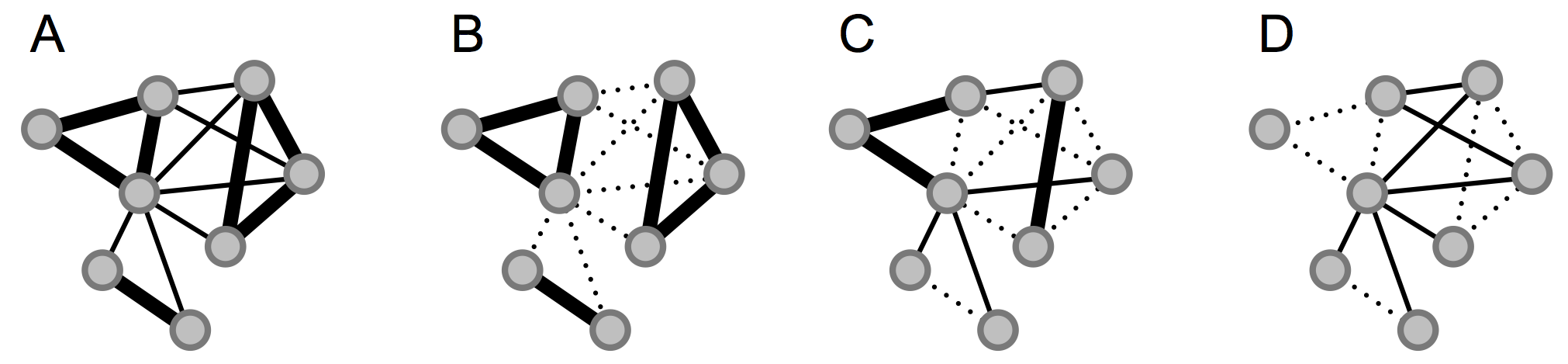}
\end{center}
\caption{ \small
{\bf Networks.} \textbf{A:} Raw network; shows all observed links for a specific time-bin. Thickness of a link symbolizes the maximum of the received signal strengths. \textbf{B:} Thresholded network, we remove links with received signal strengths below a certain threshold, where dotted lines indicate the removed links. \textbf{C:} Null model; with respect to the previous network we remove the same amount of links, but where the links are chosen at random. \textbf{D:} Control network, a similar amount of links with signal strength above or equal to the threshold are removed.
}
\label{fig:models}
\end{figure}

This section outlines various strategies for removing non-social links from the network. Fig. \ref{fig:models}A shows an illustration of the raw proximity data for a single time-bin, a link is drawn if either $i\rightarrow j$ or $j\rightarrow i$. Thickness of a link represents the strength of the received signal. For the thresholded network (Fig. \ref{fig:models}B) we remove links according to the strength of the signal (where we assume the weaker the signal the greater the relative distance between two persons). To estimate the effect of the threshold we compare it to a null model, where we remove the same number of links, but where the links are chosen at random, illustrated in Fig. \ref{fig:models}C. To minimize any noise the random removal might cause, we repeat the procedure $n=100$ times, each time choosing a new set of random links, with statistics averaged over the 100 repetitions. As a reference, to check whether thresholding actually emphasizes social proximity links, we additionally compare it to a control network, where we remove the same amount of links, but where the links have signal strengths \emph{above} or \emph{equal} to the threshold, Fig. \ref{fig:models}D. This procedure is also repeated $n$ times. In a situation where there are more links below the threshold than above, we will remove fewer links for the latter compared to the other networks.

\section*{Results}
\subsection*{Network properties}
\begin{figure}[!htbp]
\begin{center}
\includegraphics[width=0.8\linewidth]{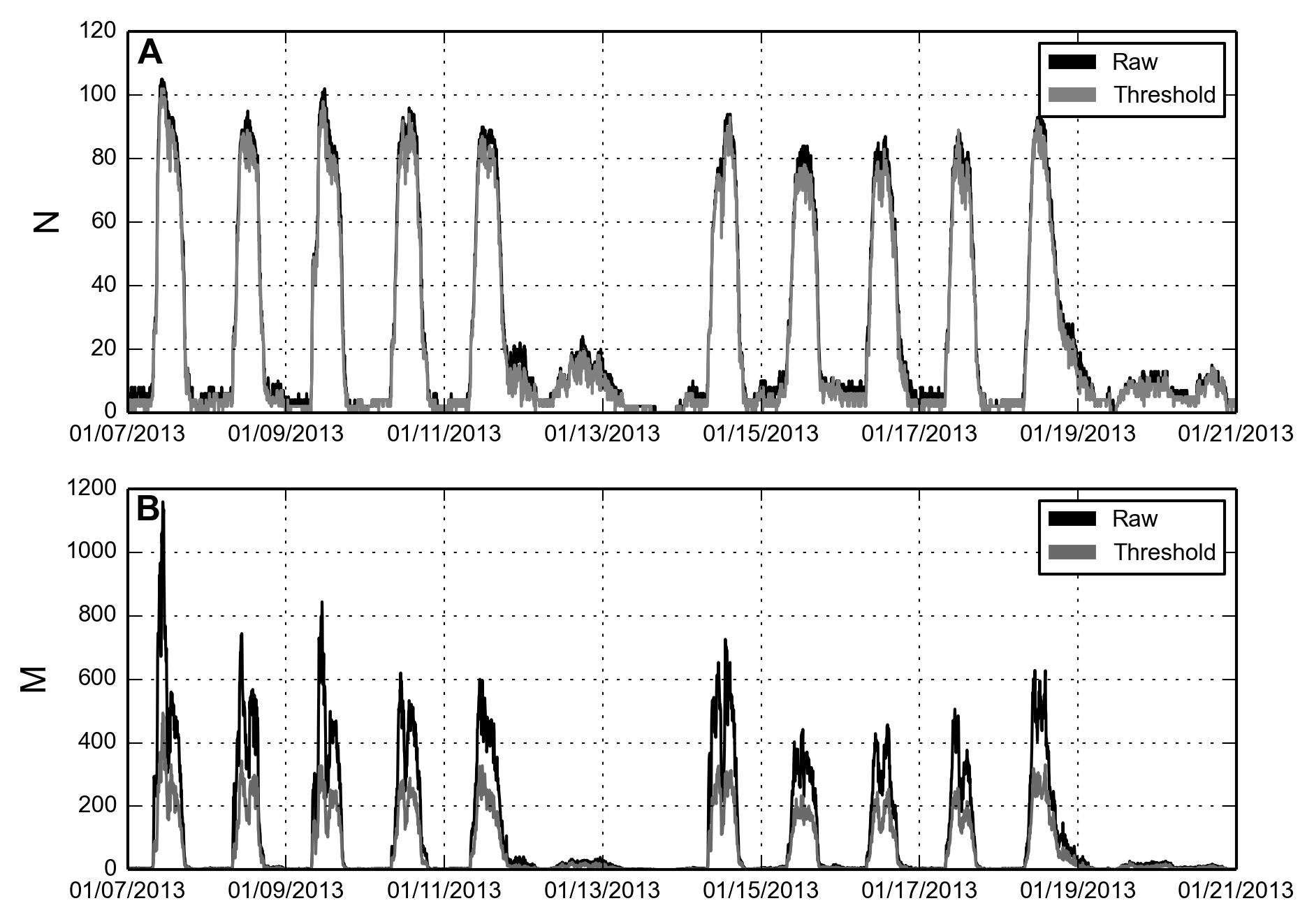}
\end{center}
\caption{ \small
{\bf Network statistics.} Properties are highly dynamic but on average we observe $17.32$ nodes and $62.50$ links per time-bin. \textbf{A:} Number of nodes $N$ as a function of time. Only active nodes are counted, i.e. people that have observed another person or been observed themselves. Dynamics are shown for two weeks during the 2013 spring semester, clearly depicting both daily and weekly patterns. Data markers are omitted to avoid visual clutter. On average thresholding removes $3.06$ nodes during weekends and holidays, and $2.38$ during regular weekdays. \textbf{B:} Number of links $M$ as a function of time. $10.60$ links are on average removed during weekends/holidays, and $32.21$ are removed during weekdays.
}
\label{fig:threshold}
\end{figure}

\begin{figure}[!htbp]
\begin{center}
\includegraphics[width=0.8\linewidth]{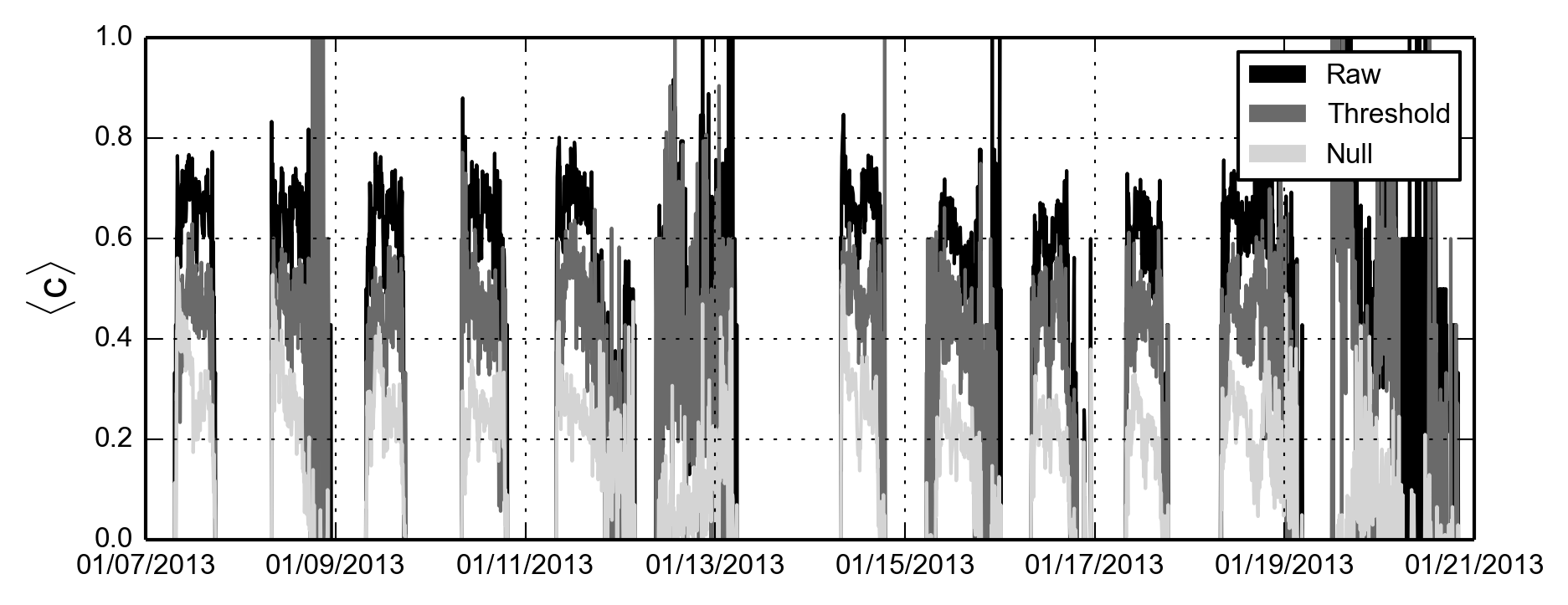}
\end{center}
\caption{ \small
{\bf Average clustering.} Only active nodes, i.e. nodes that are part of at least one dyad contribute to the average, the rest are disregarded. Average clustering is calculated according to the definition in \cite{watts1998collective}. Since social activity in groups larger than two individuals results in network triangles, the fact that clustering is not significantly reduced by thresholding (compared to the null model) provides evidence that we are preserving social structure in spite of link removal.
}
\label{fig:clust}
\end{figure}

Now that we have determined a threshold for filtering out non-social proximity links, let us study the effects on the network properties. Thresholding weak links does not significantly influence the number of nodes present ($N$) in the network (Fig. \ref{fig:threshold}A), while the number of links ($M$) is substantially reduced (Fig. \ref{fig:threshold}B). On average we remove $2.38$ nodes and $32.18$ links per time-bin.  
Social networks differ topologically from other kinds of networks by having a larger than expected number of triangles \cite{newman2003social}, thus clustering is a key component in determining the effects of thresholding. Fig. \ref{fig:clust} suggests that we are, in fact, keeping real social interactions: random removal disentangles the network and dramatically decreases the clustering coefficient, while thresholding conserves most of the average clustering. 
Calculating the average ratio ($\langle \langle c_T \rangle / \langle c_N \rangle \rangle$) between clustering in the thresholded ($\langle c_T \rangle$) and the null networks ($\langle c_N \rangle$) reveals that $c_T$ on average is $2.38$ larger. 
These findings emphasize that a selection process based on signal strength greatly differs from a random one.

\subsection*{Link evaluation}
Sorting links by signal strength and disregarding weak ones greatly reduces the number of links, but do we remove the correct links, i.e. do we get rid of noisy, non-social links? The fact that clustering remains high in spite of removing a large fraction of links is a good sign, but we want to investigate this question more directly. 
To do so, we divide the problem into two timescales; a short one where we consider the probability that a removed link might reappear a few time-steps later, and a long where we evaluate the quality of a removed link according to certain network properties. 
Let's first consider the short time-scale. We assume that human interactions take place on a time-scale that is mostly longer than the 5-minute time-bins we analyze here.
Thus, if a noisy link is removed, the probability that it will re-appear in one of the immediately following time-steps should be low, since no interaction is assumed to take place. Howbeit we expect the probability to be significantly greater than zero, since even weak (non-social) links imply physical proximity.
Similarly, if we (accidentally) remove a social link, the probability that it will appear again should be high, since the social activity is expected to continue to take place.

\begin{figure}[!htbp]
\begin{center}
\includegraphics[width=0.45\linewidth]{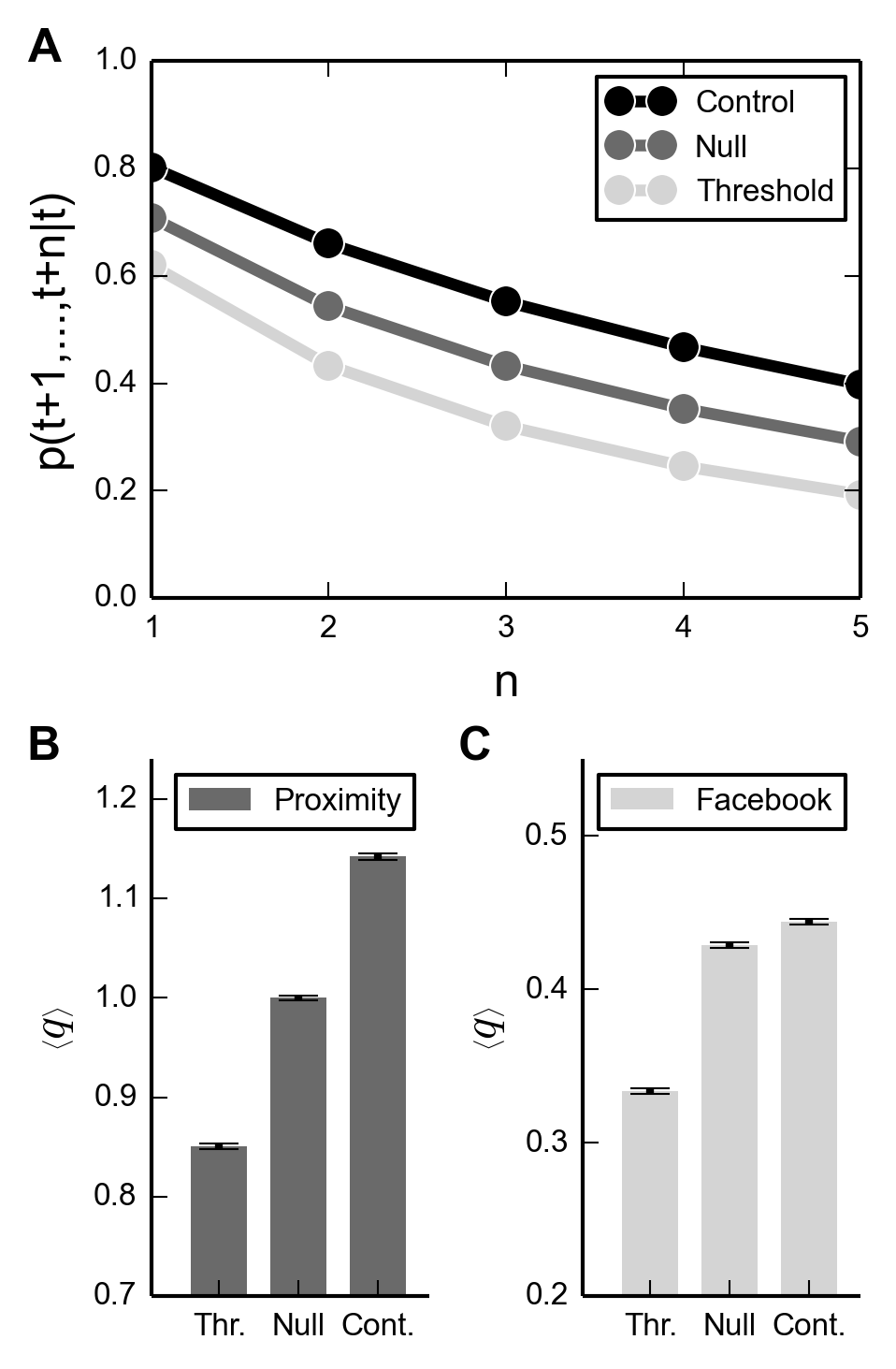}
\end{center}
\caption{ \small
{\bf Link evaluation.} \textbf{A:} Probability of link reappearance. For each selection process we remove a specific set of links. In the thresholded network, we remove links with weak signal strength. For the null network, we remove links at random. Lastly, in the control network case we remove strong links. The probability for links to reappear within all the next $n$ time-steps is calculated using Eq. \ref{eq:reappear} and averaging over all time-bins. Boundary conditions are not applied and the reappearance probability for the last $n=5$ bins is not taken into account. \textbf{B:} Quality measure for proximity data. \textbf{C:} Quality measure for the online data. For each time-bin we calculate $q_t$ as defined in Eq. \ref{eq:proximity} and \ref{eq:online}. Brackets indicate a temporal average across all time-bins and value are shown for all three network types.
}
\label{fig:link_evaluation}
\end{figure}

Let us formalize this notion. Consider a link $e$ that is removed at time $t$, the probability that the link will appear in the next time-step is $p(t+1|e,t)$. Generalizing this we can write the probability that any removed link will appear in all the following $n$ time-steps as:
\begin{equation}
p(t+1,\ldots,t+n|t)=\frac{\text{no. links removed at } t \text{ present at } t+1 \cap \ldots \cap t+n}{\text{no. links removed at } t}
\label{eq:reappear}
\end{equation}
Fig.~\ref{fig:link_evaluation}A illustrates that thresholded links in subsequent time-steps are observed less frequently then both null and control links. To compare with the worst possible condition, we compare data from each thresholded time-bin with the \emph{raw data} from the next bin (where the raw data contains many weak links). In spite of this, we observe a clear advantage of distinguishing between links with weak and strong signal strengths. If we look at values for $t+1$, the first subsequent time-step, the probability of re-occurrence in the thresholded network is about $12\%$ lower than for the null model, and as we look to later time-steps, the gap widens.

A different set of social dynamics unfolds on longer timescales where the class schedule imposes certain links to appear periodically, e.g every week. Here we determine impact of removing links in two ways. First, we use total link weights and second, we use online friendship status. Friends meet frequently; we capture this behavior by using the total number of observations of a certain dyad to estimate the weight of a friendship (again, counted in the raw network). Thus, we evaluate the quality of a removed links by considering its total weight compared to the weight of other links present in the same time-bin. However, since multiple links are removed per time-bin we are more interested in the average,
\begin{equation} \label{eq:proximity}
q_t=\frac{\text{Avg. weight of removed links at } t}{\text{Avg. weight of all links present at } t}
\end{equation}
This estimates, per time-bin, whether removed links on average have weights below, close to, or above the mean.
Note that the measure is intended to estimate the quality of removed links and is therefore not defined for bins where zero links are removed.
Fig. \ref{fig:link_evaluation}B indicates difference in link selection processes.
Choosing links at random (null network) removes both strong and weak links with equal probability, thus on average this corresponds to the mean weight of links present.
Compared to null, the thresholded network removes links with weights below average, indicating that removed links are less frequently observed and therefore also less likely to be real friendships.
The control case displays an diametrical behavior, on average, it removes links with higher weights. 

The second method to evaluate the link-selection processes compares the set of removed links with the structure of an online social network, i.e. if a removed proximity link has an equivalent online counterpart. 
We estimate the quality by measuring the fraction of removed links with respect to those present at time $t$.
\begin{equation} \label{eq:online}
q_t^{\text{FB}}=\frac{\text{no. of FB links removed at } t }{\text{no. of FB links present at } t}
\end{equation}
The quality measure is essentially a ratio, i.e.~it can assume values $0 \leq q_t^{\text{FB}} \leq 1$ depending on the fraction of links that are removed.
Bins with zero Facebook friendships are disregarded since they contain no information regarding the online social network.
Fig.~\ref{fig:link_evaluation}C shows that random removal (null network), on average, removes $\sim 43 \%$ of online friendships, while the thresholded network removes $\sim 33 \%$, a $10$ percent point difference.
For comparison, the control network removes $\sim 44\%$ of the online links.
Further, redoing the analysis for a dataset comprised only of users for which we have both proximity and online data for, does not significantly alter the results.

Facebook links are not necessary good indicators for strong friendships, but are more likely to correspond to real social interactions.
In spite of this, both Fig. \ref{fig:link_evaluation}B and C support that distinguishing between strong and weak proximity links tends to emphasize real social interactions: on average thresholded links have lower edge weights and remove fewer Facebook friendships compared to both the null-model and the control.

\section*{Discussion}
The availability of electronic datasets is increasing, so the question of how well can we use these electronic \textit{clicks} to infer actual social interactions is important for effectively understanding processes such as relational dynamics, and contagion. 
Sorting links based on their signal strength allows us to distinguish between strong and weak ties, and we have argued that thresholding the network emphasizes social proximity links while eliminating some noise. 

Simply thresholding links based on signal strength is not a perfect solution. In certain settings we remove real social connections while noisy links are retained. Our results indicate that the proposed framework is better at identifying strong links than removing them. A trend which the link-reappearance probability, link-weights, and online friendship analysis support. Compared to the baseline we achieve better results than just assuming all proximity observations as real social interactions.
But determining whether a close proximity link corresponds to an actual friendship interaction is much more difficult. Multiple scenarios exist where people are in close contact but are not friends, one obvious example is queuing.
Each human interaction has a specific social context, so an understanding of the underlying social fabric is required to fully discern when a close proximity link is an actual social meeting. This brings us back to the question of how to determine a real friendship from digital observations (cf.~\cite{wuchty2009social}). Close proximity may not be the best indicator of friendship; call logs, text logs, and geographical positions are all factors which coupled with information from the Bluetooth probe could give us a better insight into social dynamics and interactions. 

\section*{Acknowledgments}
We thank L. K. Hansen, A. Stopczynski, and P. Sapie\.{z}y\'{n}ski for many useful discussions and A. Cuttone for proofreading the manuscript.
The work in this paper was funded by a Young Investigator Grant from the Villum Foundation (High Resolution Networks, awarded to SL).

\newpage
\bibliography{bluetooth}

\begin{thebibliography}{10}
\providecommand{\url}[1]{\texttt{#1}}
\providecommand{\urlprefix}{URL }
\expandafter\ifx\csname urlstyle\endcsname\relax
  \providecommand{\doi}[1]{doi:\discretionary{}{}{}#1}\else
  \providecommand{\doi}{doi:\discretionary{}{}{}\begingroup
  \urlstyle{rm}\Url}\fi
\providecommand{\bibAnnoteFile}[1]{%
  \IfFileExists{#1}{\begin{quotation}\noindent\textsc{Key:} #1\\
  \textsc{Annotation:}\ \input{#1}\end{quotation}}{}}
\providecommand{\bibAnnote}[2]{%
  \begin{quotation}\noindent\textsc{Key:} #1\\
  \textsc{Annotation:}\ #2\end{quotation}}
\providecommand{\eprint}[2][]{\url{#2}}

\bibitem{sun2011exploring}
Sun J, Yuan J, Wang Y, Si H, Shan X (2011) Exploring space--time structure of
  human mobility in urban space.
\newblock Physica A: Statistical Mechanics and its Applications 390: 929--942.
\bibAnnoteFile{sun2011exploring}

\bibitem{sevtsuk2010does}
Sevtsuk A, Ratti C (2010) Does urban mobility have a daily routine? learning
  from the aggregate data of mobile networks.
\newblock Journal of Urban Technology 17: 41--60.
\bibAnnoteFile{sevtsuk2010does}

\bibitem{liljeros2001web}
Liljeros F, Edling CR, Amaral LAN, Stanley HE, {\AA}berg Y (2001) The web of
  human sexual contacts.
\newblock Nature 411: 907--908.
\bibAnnoteFile{liljeros2001web}

\bibitem{mossong2008social}
Mossong J, Hens N, Jit M, Beutels P, Auranen K, et~al. (2008) Social contacts
  and mixing patterns relevant to the spread of infectious diseases.
\newblock {PL}o{S} {M}edicine 5: e74.
\bibAnnoteFile{mossong2008social}

\bibitem{cauchemez2009household}
Cauchemez S, Donnelly CA, Reed C, Ghani AC, Fraser C, et~al. (2009) Household
  transmission of 2009 pandemic influenza a ({H1N1}) virus in the united
  states.
\newblock New England Journal of Medicine 361: 2619--2627.
\bibAnnoteFile{cauchemez2009household}

\bibitem{wu2008mining}
Wu L, Waber B, Aral S, Brynjolfsson E, Pentland A (2008) Mining face-to-face
  interaction networks using sociometric badges: {P}redicting productivity in
  an {IT} configuration task.
\newblock Available at {SSRN} 1130251.
\bibAnnoteFile{wu2008mining}

\bibitem{pentland2012new}
Pentland A (2012) The new science of building great teams.
\newblock Harvard Business Review 90: 60--69.
\bibAnnoteFile{pentland2012new}

\bibitem{blansky2013spread}
Blansky D, Kavanaugh C, Boothroyd C, Benson B, Gallagher J, et~al. (2013)
  Spread of academic success in a high school social network.
\newblock {PL}o{S} {ONE} 8: e55944.
\bibAnnoteFile{blansky2013spread}

\bibitem{wuchty2009social}
Wuchty S (2009) What is a social tie?
\newblock Proceedings of the National Academy of Sciences 106: 15099--15100.
\bibAnnoteFile{wuchty2009social}

\bibitem{watts2007twenty}
Watts DJ (2007) A twenty-first century science.
\newblock Nature 445: 489--489.
\bibAnnoteFile{watts2007twenty}

\bibitem{eckmann2004entropy}
Eckmann JP, Moses E, Sergi D (2004) Entropy of dialogues creates coherent
  structures in e-mail traffic.
\newblock Proceedings of the National Academy of Sciences of the United States
  of America 101: 14333--14337.
\bibAnnoteFile{eckmann2004entropy}

\bibitem{barabasi2005origin}
Barabasi AL (2005) The origin of bursts and heavy tails in human dynamics.
\newblock Nature 435: 207--211.
\bibAnnoteFile{barabasi2005origin}

\bibitem{kossinets2006empirical}
Kossinets G, Watts DJ (2006) Empirical analysis of an evolving social network.
\newblock Science 311: 88--90.
\bibAnnoteFile{kossinets2006empirical}

\bibitem{onnela2007structure}
Onnela JP, Saram{\"a}ki J, Hyv{\"o}nen J, Szab{\'o} G, Lazer D, et~al. (2007)
  Structure and tie strengths in mobile communication networks.
\newblock Proceedings of the National Academy of Sciences 104: 7332--7336.
\bibAnnoteFile{onnela2007structure}

\bibitem{gonzalez2008understanding}
Gonzalez MC, Hidalgo CA, Barabasi AL (2008) Understanding individual human
  mobility patterns.
\newblock Nature 453: 779--782.
\bibAnnoteFile{gonzalez2008understanding}

\bibitem{lazer2009computational}
Lazer D, Pentland A, Adamic L, Aral S, Barab{\'a}si AL, et~al. (2009)
  Computational social science.
\newblock Science 323: 721-723.
\bibAnnoteFile{lazer2009computational}

\bibitem{song2010limits}
Song C, Qu Z, Blumm N, Barab{\'a}si AL (2010) Limits of predictability in human
  mobility.
\newblock Science 327: 1018--1021.
\bibAnnoteFile{song2010limits}

\bibitem{bagrow2011collective}
Bagrow JP, Wang D, Barab{\'a}si AL (2011) Collective response of human
  populations to large-scale emergencies.
\newblock {PL}o{S} {ONE} 6: e17680.
\bibAnnoteFile{bagrow2011collective}

\bibitem{eagle2009inferring}
Eagle N, Pentland AS, Lazer D (2009) Inferring friendship network structure by
  using mobile phone data.
\newblock Proceedings of the National Academy of Sciences 106: 15274--15278.
\bibAnnoteFile{eagle2009inferring}

\bibitem{haritaoglu2000w}
Haritaoglu I, Harwood D, Davis LS (2000) W4: Real-time surveillance of people
  and their activities.
\newblock Pattern Analysis and Machine Intelligence, IEEE Transactions on 22:
  809--830.
\bibAnnoteFile{haritaoglu2000w}

\bibitem{polastre2005telos}
Polastre J, Szewczyk R, Culler D (2005) Telos: enabling ultra-low power
  wireless research.
\newblock In: Information Processing in Sensor Networks, 2005. IPSN 2005.
  Fourth International Symposium on. IEEE, pp. 364--369.
\bibAnnoteFile{polastre2005telos}

\bibitem{salathe2010high}
Salath{\'e} M, Kazandjieva M, Lee JW, Levis P, Feldman MW, et~al. (2010) A
  high-resolution human contact network for infectious disease transmission.
\newblock Proceedings of the National Academy of Sciences 107: 22020--22025.
\bibAnnoteFile{salathe2010high}

\bibitem{rosenstein2008video}
Rosenstein B (2008) Video use in social science research and program
  evaluation.
\newblock International Journal of Qualitative Methods 1: 22--43.
\bibAnnoteFile{rosenstein2008video}

\bibitem{olguin2009sensible}
Olgu{\'\i}n DO, Waber BN, Kim T, Mohan A, Ara K, et~al. (2009) Sensible
  organizations: Technology and methodology for automatically measuring
  organizational behavior.
\newblock Systems, Man, and Cybernetics, Part B: Cybernetics, IEEE Transactions
  on 39: 43--55.
\bibAnnoteFile{olguin2009sensible}

\bibitem{kjaergaard2012challenges}
Kj{\ae}rgaard MB, Nurmi P (2012) Challenges for social sensing using wifi
  signals.
\newblock In: Proceedings of the 1st ACM workshop on Mobile systems for
  computational social science. ACM, pp. 17--21.
\bibAnnoteFile{kjaergaard2012challenges}

\bibitem{wyatt2007capturing}
Wyatt D, Choudhury T, Kautz H (2007) Capturing spontaneous conversation and
  social dynamics: A privacy-sensitive data collection effort.
\newblock In: Acoustics, Speech and Signal Processing, 2007. ICASSP 2007. IEEE
  International Conference on. IEEE, volume~4, pp. IV--213.
\bibAnnoteFile{wyatt2007capturing}

\bibitem{carreras2012comm2sense}
Carreras I, Matic A, Saar P, Osmani V (2012) Comm2sense: Detecting proximity
  through smartphones.
\newblock In: Pervasive Computing and Communications Workshops (PERCOM
  Workshops), 2012 IEEE International Conference on. IEEE, pp. 253--258.
\bibAnnoteFile{carreras2012comm2sense}

\bibitem{wang2011human}
Wang D, Pedreschi D, Song C, Giannotti F, Barabasi AL (2011) Human mobility,
  social ties, and link prediction.
\newblock In: Proceedings of the 17th ACM SIGKDD international conference on
  Knowledge discovery and data mining. ACM, pp. 1100--1108.
\bibAnnoteFile{wang2011human}

\bibitem{cattuto2010dynamics}
Cattuto C, Van~den Broeck W, Barrat A, Colizza V, Pinton JF, et~al. (2010)
  Dynamics of person-to-person interactions from distributed {RFID} sensor
  networks.
\newblock {PL}o{S} {ONE} 5: e11596.
\bibAnnoteFile{cattuto2010dynamics}

\bibitem{barrat2013temporal}
Barrat A, Cattuto C (2013) Temporal networks of face-to-face human
  interactions.
\newblock In: Temporal Networks, Springer. pp. 191--216.
\bibAnnoteFile{barrat2013temporal}

\bibitem{stopczynski2014measuring}
Stopczynski A, Sekara V, Sapiezynski P, Cuttone A, Madsen MM, et~al. (2014)
  Measuring large-scale social networks with high resolution.
\newblock {PL}o{S} {ONE} 9: e95978.
\bibAnnoteFile{stopczynski2014measuring}

\bibitem{ladd2005robotics}
Ladd AM, Bekris KE, Rudys A, Kavraki LE, Wallach DS (2005) Robotics-based
  location sensing using wireless ethernet.
\newblock Wireless Networks 11: 189--204.
\bibAnnoteFile{ladd2005robotics}

\bibitem{shorey2000bluetooth}
Shorey R, Miller BA (2000) The bluetooth technology: merits and limitations.
\newblock In: Personal Wireless Communications, 2000 IEEE International
  Conference on. IEEE, pp. 80--84.
\bibAnnoteFile{shorey2000bluetooth}

\bibitem{clauset2012persistence}
Clauset A, Eagle N (2012) Persistence and periodicity in a dynamic proximity
  network.
\newblock arXiv preprint arXiv:12117343 .
\bibAnnoteFile{clauset2012persistence}

\bibitem{sulo2010meaningful}
Sulo R, Berger-Wolf T, Grossman R (2010) Meaningful selection of temporal
  resolution for dynamic networks.
\newblock In: Proceedings of the Eighth Workshop on Mining and Learning with
  Graphs. ACM, pp. 127--136.
\bibAnnoteFile{sulo2010meaningful}

\bibitem{cheung2006inexpensive}
Cheung KC, Intille SS, Larson K (2006) An inexpensive bluetooth-based indoor
  positioning hack.
\newblock Proc UbiComp06 Extended Abstracts .
\bibAnnoteFile{cheung2006inexpensive}

\bibitem{anastasi2003experimenting}
Anastasi G, Bandelloni R, Conti M, Delmastro F, Gregori E, et~al. (2003)
  Experimenting an indoor bluetooth-based positioning service.
\newblock In: Distributed Computing Systems Workshops, 2003. Proceedings. 23rd
  International Conference on. IEEE, pp. 480--483.
\bibAnnoteFile{anastasi2003experimenting}

\bibitem{bruno2003design}
Bruno R, Delmastro F (2003) Design and analysis of a bluetooth-based indoor
  localization system.
\newblock In: Personal wireless communications. Springer, pp. 711--725.
\bibAnnoteFile{bruno2003design}

\bibitem{madhavapeddy2005study}
Madhavapeddy A, Tse A (2005) A study of bluetooth propagation using accurate
  indoor location mapping.
\newblock In: UbiComp 2005: Ubiquitous Computing, Springer. pp. 105--122.
\bibAnnoteFile{madhavapeddy2005study}

\bibitem{zhou2006position}
Zhou S, Pollard JK (2006) Position measurement using bluetooth.
\newblock Consumer Electronics, IEEE Transactions on 52: 555--558.
\bibAnnoteFile{zhou2006position}

\bibitem{hay2009bluetooth}
Hay S, Harle R (2009) Bluetooth tracking without discoverability.
\newblock In: Location and context awareness, Springer. pp. 120--137.
\bibAnnoteFile{hay2009bluetooth}

\bibitem{hossain2007comprehensive}
Hossain A, Soh WS (2007) A comprehensive study of bluetooth signal parameters
  for localization.
\newblock In: Personal, Indoor and Mobile Radio Communications, 2007. PIMRC
  2007. IEEE 18th International Symposium on. IEEE, pp. 1--5.
\bibAnnoteFile{hossain2007comprehensive}

\bibitem{friis1946note}
Friis HT (1946) A note on a simple transmission formula.
\newblock proc IRE 34: 254--256.
\bibAnnoteFile{friis1946note}

\bibitem{liu2011accurate}
Liu S, Striegel A (2011) Accurate extraction of face-to-face proximity using
  smartphones and bluetooth.
\newblock In: Computer Communications and Networks (ICCCN), 2011 Proceedings of
  20th International Conference on. IEEE, pp. 1--5.
\bibAnnoteFile{liu2011accurate}

\bibitem{hall1990hidden}
Hall ET (1990) The hidden dimension.
\newblock Anchor Books New York.
\bibAnnoteFile{hall1990hidden}

\bibitem{newman2003social}
Newman ME, Park J (2003) Why social networks are different from other types of
  networks.
\newblock Physical Review E 68: 036122.
\bibAnnoteFile{newman2003social}

\bibitem{watts1998collective}
Watts DJ, Strogatz SH (1998) Collective dynamics of 'small-world' networks.
\newblock Nature 393: 440--442.
\bibAnnoteFile{watts1998collective}

\end{thebibliography}

\end{document}